# A Novel Generic Session Based Bit Level Encryption Technique to Enhance Information Security

Manas Paul,[1] Tanmay Bhattacharya,[2]
[1]Sr. Lecturer, Department of Computer Application, JIS College of Engineering, Kalyani, West Bengal
e-mail:manaspaul@rediffmail.com
[2]Sr. Lecturer, Department of Information Technology, JIS College of Engineering, Kalyani, West Bengal
e-mail: tanmay_bhattacharya@yahoo.co.in

Suvajit Pal,[3] Ranit Saha[4]
[3]Student, Department of Information Technology, JIS College of Engineering, Kalyani, West Bengal
e-mail:suvajitpal@gmail.com
[4] Student, Department of Information Technology, JIS College of Engineering, Kalyani, West Bengal
e-mail:ranit.saha@gmail.com

*Abstract* - **In this paper a session based symmetric key encryption system has been proposed and is termed as Permutated Cipher Technique (PCT). This technique is more fast, suitable and secure for larger files. In this technique the input file is broken down into blocks of various sizes (of 2^n order) and encrypted by shifting the position of each bit by a certain value for a certain number of times. A key is generated randomly wherein the length of each block is determined. Each block length generates a unique value of "number of bits to be skipped". This value determines the new position of the bits within the block that are to be shifted. After the shifting and inverting each block is XOR'ed with SHA-512 digest of the key. The resultant blocks from the cipher text. The key is generated according to the binary value of the input file size.**
   **Decryption is done following the same process as the technique is symmetric.**

## I. INTRODUCTION

   In the age of science and technology everybody is using internet in almost every discipline of their daily life. Internet has made our communication faster and easier. Hence, maintaining the security of essential information is of utmost importance. Therefore many researchers are working in the field of encrypting the data transacted through internet. Encryption process converts the transacted data into cipher text which is otherwise illegible without the proper decrypting technique. Many algorithms based on various mathematical models are available, but each of them has their own share of merits and demerits. No single algorithm is perfect. As a result, continual researches are being made in the field of cryptography to enhance the security even further.

   In this paper a new technique has been proposed, where the source message is considered as a stream of binary bits whose positions are shifted [1] to create the encrypted message. The key is generated in a unique random manner [2, 3] (cryptographically strong pseudo-random number generator RFC 1750) and all the variables involved are calculated on the basis of the random generated sequence of the numbers within the key. Section II of the paper discusses about the proposed scheme with block diagrams. Section III discussed about the key generation, the encryption and decryption procedure, section IV shows the results on different files and comparison of the proposed technique with TDES [4] and AES [5], section V & VI deals with conclusions & future scope respectively.

## II. THE SCHEME

The PCT algorithm consists of three major divisions:
- Key Generation
- Encryption Mechanism
- Decryption Mechanism

*Key Generation:*

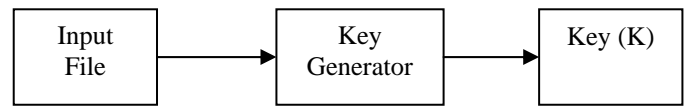

*Encryption Mechanism:*

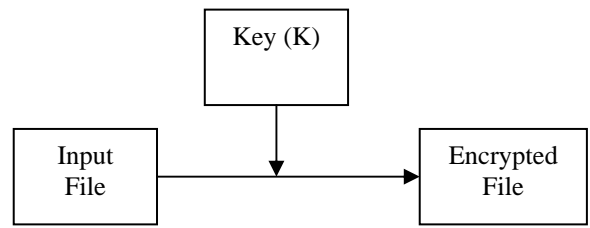

*Decryption Mechanism:*

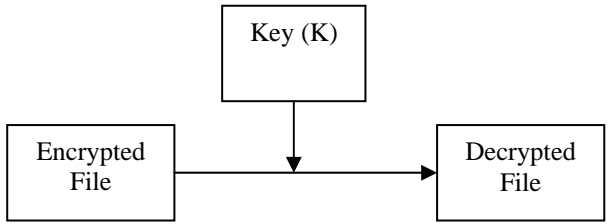

## III. PROPOSED ALGORITHM

**Key Generation Algorithm:**

1. Total length of the input file is calculated and the corresponding binary representation of the above value is stored in an array.
2. Two index values of the array (mH, mL) are selected randomly.
3. The value of the content of mH cell be decreased by a certain number "x" and "x" is multiplied to $2^{(mH-mL)}$ and added to the contents of mL cell. The steps are repeated for random number of times.
4. The index values are written into another array, repeating them the number of times of their cell value.
5. The contents of this new array are swapped for random number of times.

Example:

| 8 | 7 | 6 | 5 | 4 | 3 | 2 | 1 |
|---|---|---|---|---|---|---|---|
| 1 | 1 | 0 | 1 | 1 | 1 | 0 | 1 |

| 8 | 7 | 6 | **5** | 4 | 3 | 2 | **1** |
|---|---|---|---|---|---|---|---|
| 1 | 1 | 0 | 1 | 1 | 1 | 0 | 1 |

mH = 5; mL = 1 : a [mH] = 1-1 = 0;
a [mL] = 1 x 2 ^ (mH-mL) + a [mL]
a [mL] = 1 x 2 ^ (5-1) + 1 = 17

| 8 | 7 | 6 | **5** | 4 | 3 | 2 | **1** |
|---|---|---|---|---|---|---|---|
| 1 | 1 | 0 | 0 | 1 | 1 | 0 | 0 |

After repeating random no. of times we have something like:

| 8 | 7 | 6 | 5 | 4 | 3 | 2 | 1 |
|---|---|---|---|---|---|---|---|
| 3 | 5 | 4 | 17 | 8 | 9 | 11 | 12 |

New Array (writing the index values):

| 8 | 7 | 6 | 5 | 4 | 3 | 2 | 1 |
|---|---|---|---|---|---|---|---|
| 1 | 1 | 1 | 2 | 2 | 2 | 2 | 2 |

…..
Swapped New Array gives a random sequence:

| 8 | 7 | 6 | 5 | 4 | 3 | 2 | 1 |
|---|---|---|---|---|---|---|---|
| 2 | 4 | 2 | 1 | 1 | 5 | 4 | 2 |

…..

**Encryption Algorithm:**

1. A block size is read from key. If block size is sufficed from available byte input stream then proceed, else required padding be added to achieve block size.
2. Odd bits are flipped.
3. Now Consider bits only in even positions, let i denote the even position containing a bit.
    For each even bit:
    - ith position bit from encrypted block is extracted
    - The bit is set to the correct position. The bit is set using the bitwise OR | operator.
    - i is increased by number of bits to skip (NBSk).
4. Finally after the entire block in encrypted, we XOR with SHA-512 [6] digest of the key.

**Decryption Algorithm:**

1. A block size is read from key. The necessary variables like MaxIter is calculated

2. Again XOR ed with SHA-512 digest of the key.

3. Bits only in even positions are considered, if i denote the even position containing a bit.

Then for each even bit:
- ith position bit from encrypted block is extracted
- The bit is set to the correct position. The bit is set using the bitwise OR | operator.
- i is increased by number of bits to skip (NBSk).

4. Odd bits are fliped again to get back original bits in odd position

## IV. RESULTS AND ANALYSIS

In this section the implementation of different types of files are presented. Files are chosen at random comprising of various file sizes and file types. Here we present analysis of 20 files of 8 different file types, varying file sizes from 330 bytes to 62657918 bytes (59.7MB) on three standard algorithms viz. Triple-DES 168bit, AES 128bit; and the proposed PCT algorithm. Analysis includes comparing encryption and decryption times, Chi-Square values, Avalanche and Strict Avalanche effects and Bit Independence. Implementation of all algorithms and different types of tests has been done using JAVA.

### IV-I. ENCRYPTION AND DECRYPTION TIME

Table I & Table II shows the encryption and decryption time against increasing size of files for Triple-DES 168bit, AES 128bit and the proposed PCT techniques. For most of the file size and file types the proposed PCT takes less time to encrypt/decrypt compared to T-DES and nearly same time for decryption compared to AES technique. From table it's seen that as the file size increases the proposed PCT performs much better than Triple-DES and its performance matches the AES technique. For even larger file, of size greater than 1GB the proposed PCT technique can even outperform AES. Fig. 1and 2 shows the graphical representation of the same in logarithmic scale.

Table I
File size v/s encryption time
(for Triple-DES, AES and PCT algorithms)

|  | Source File Size (in bytes) | File type | Encryption Time (in seconds) | | |
|---|---|---|---|---|---|
|  |  |  | TDES | AES | PCT |
| 1 | 330 | dll | 0.001 | 0.001 | 0.004 |
| 2 | 528 | txt | 0.001 | 0.001 | 0.008 |
| 3 | 96317 | txt | 0.034 | 0.004 | 0.020 |
| 4 | 233071 | rar | 0.082 | 0.011 | 0.062 |
| 5 | 354304 | exe | 0.123 | 0.017 | 0.081 |
| 6 | 536387 | zip | 0.186 | 0.023 | 0.133 |
| 7 | 657408 | doc | 0.220 | 0.031 | 0.234 |
| 8 | 682496 | dll | 0.248 | 0.031 | 0.066 |
| 9 | 860713 | pdf | 0.289 | 0.038 | 0.114 |
| 10 | 988216 | exe | 0.331 | 0.042 | 0.155 |
| 11 | 1395473 | txt | 0.476 | 0.059 | 0.165 |
| 12 | 4472320 | doc | 1.663 | 0.192 | 0.371 |
| 13 | 7820026 | avi | 2.626 | 0.334 | 0.651 |
| 14 | 9227808 | zip | 3.096 | 0.397 | 0.474 |
| 15 | 11580416 | dll | 4.393 | 0.544 | 0.792 |
| 16 | 17486968 | exe | 5.906 | 0.743 | 1.884 |
| 17 | 20951837 | rar | 7.334 | 0.937 | 1.578 |
| 18 | 32683952 | pdf | 10.971 | 1.350 | 2.077 |
| 19 | 44814336 | exe | 15.091 | 1.914 | 2.974 |
| 20 | 62657918 | avi | 21.133 | 2.689 | 5.870 |

Table II
File size v/s decryption time
(for Triple-DES, AES and PCT algorithms)

|  | Source File Size (in bytes) | File type | Decryption Time (in seconds) | | |
|---|---|---|---|---|---|
|  |  |  | TDES | AES | PCT |
| 1 | 330 | dll | 0.001 | 0.001 | 0.002 |
| 2 | 528 | txt | 0.001 | 0.001 | 0.006 |
| 3 | 96317 | txt | 0.035 | 0.008 | 0.027 |
| 4 | 233071 | rar | 0.087 | 0.017 | 0.056 |
| 5 | 354304 | exe | 0.128 | 0.025 | 0.069 |
| 6 | 536387 | zip | 0.202 | 0.038 | 0.058 |
| 7 | 657408 | doc | 0.235 | 0.045 | 0.198 |
| 8 | 682496 | dll | 0.266 | 0.046 | 0.128 |
| 9 | 860713 | pdf | 0.307 | 0.060 | 0.088 |
| 10 | 988216 | exe | 0.356 | 0.070 | 0.130 |
| 11 | 1395473 | txt | 0.530 | 0.098 | 0.298 |
| 12 | 4472320 | doc | 1.663 | 0.349 | 0.482 |
| 13 | 7820026 | avi | 2.832 | 0.557 | 0.594 |
| 14 | 9227808 | zip | 3.377 | 0.656 | 0.448 |
| 15 | 11580416 | dll | 4.652 | 0.868 | 0.871 |
| 16 | 17486968 | exe | 6.289 | 1.220 | 1.575 |
| 17 | 20951837 | rar | 8.052 | 1.431 | 1.803 |
| 18 | 32683952 | pdf | 11.811 | 2.274 | 3.312 |
| 19 | 44814336 | exe | 16.253 | 3.108 | 2.948 |
| 20 | 62657918 | avi | 22.882 | 4.927 | 5.300 |

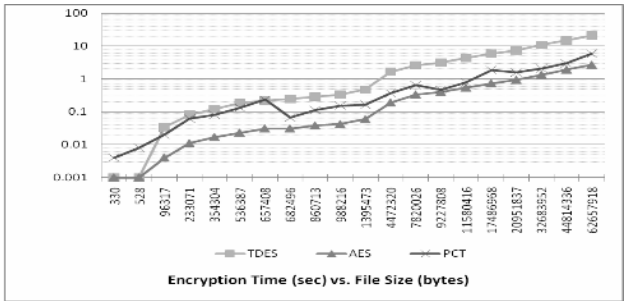

Fig. 1. Encryption Time (sec) vs. File Size (bytes) in logarithmic scale

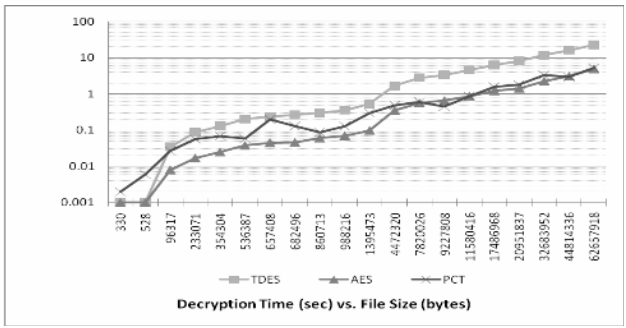

Fig. 2. Decryption Time (sec) vs. File Size (bytes) in logarithmic scale

IV-II. STUDIES ON AVALANCHE, STRICT AVALANCHE EFFECTS AND BIT INDEPENDENCE CRITERION.

Avalanche, Strict Avalanche effects and Bt Independence criterion are measured using statistical analysis of data. The bit changes among encrypted bytes for a single bit change in the original message sequence for the entire or a relative large number of bytes. The Standard Deviation from the expected values is calculated. We subtract the ratio of calculated standard deviation with expected value from 1.0 to get the Avalanche and Strict Avalanche achieve effect on a 0.0 - 1.0 scale. The closer the value is to 1.0, we achieve better Avalanche, Strict Avalanche effects and Bit Independence criterion. We take into consideration up to 5 significant digits to the right of the decimal point for more accurate interpretation. For better visual interpretation from the graphs the y-axis scale for Avalanche and Strict Avalanche effects are from 0.9 to 1.0 Table III, Table IV and Table V shows the Avalanche, Strict Avalanche and Bit Independence respectively. Fig. 3, 4 and 5 shows the graphical representation of the same.

Table III
Avalanche effect for TDES, AES and PCT algorithms.

|   | Source File Size (in bytes) | File type | Avalanche achieved | | |
|---|---|---|---|---|---|
|   |   |   | TDES | AES | PCT |
| 1 | 330 | dll | 0.99591 | 0.98904 | 0.96537 |
| 2 | 528 | txt | 0.99773 | 0.99852 | 0.97761 |
| 3 | 96317 | txt | 0.99996 | 0.99997 | 0.99083 |
| 4 | 233071 | rar | 0.99994 | 0.99997 | 0.99558 |
| 5 | 354304 | exe | 0.99996 | 0.99999 | 0.99201 |
| 6 | 536387 | zip | 0.99996 | 0.99994 | 0.99671 |
| 7 | 657408 | doc | 0.99996 | 0.99999 | 0.99438 |
| 8 | 682496 | dll | 0.99998 | 1.00000 | 0.99285 |
| 9 | 860713 | pdf | 0.99996 | 0.99997 | 0.99485 |
| 10 | 988216 | exe | 1.00000 | 0.99998 | 0.98785 |
| 11 | 1395473 | txt | 1.00000 | 1.00000 | 0.99452 |
| 12 | 4472320 | doc | 0.99999 | 0.99997 | 0.99100 |
| 13 | 7820026 | avi | 1.00000 | 0.99999 | 0.99555 |
| 14 | 9227808 | zip | 1.00000 | 1.00000 | 0.99999 |
| 15 | 11580416 | dll | 1.00000 | 0.99999 | 0.99865 |
| 16 | 17486968 | exe | 1.00000 | 0.99999 | 0.99908 |
| 17 | 20951837 | rar | 1.00000 | 1.00000 | 1.00000 |
| 18 | 32683952 | pdf | 0.99999 | 1.00000 | 0.99996 |
| 19 | 44814336 | exe | 0.99997 | 0.99997 | 0.99986 |
| 20 | 62657918 | avi | 0.99999 | 0.99999 | 0.99994 |

Table IV
Strict Avalanche effect for TDES, AES & PCT algorithms.

|   | Source File Size (bytes) | File type | Strict Avalanche achieved | | |
|---|---|---|---|---|---|
|   |   |   | TDES | AES | PCT |
| 1 | 330 | dll | 0.98645 | 0.98505 | 0.89638 |
| 2 | 528 | txt | 0.99419 | 0.99311 | 0.96829 |
| 3 | 96317 | txt | 0.99992 | 0.99987 | 0.97723 |
| 4 | 233071 | rar | 0.99986 | 0.99985 | 0.99236 |
| 5 | 354304 | exe | 0.99991 | 0.99981 | 0.98872 |
| 6 | 536387 | zip | 0.99988 | 0.99985 | 0.99552 |
| 7 | 657408 | doc | 0.99989 | 0.99990 | 0.99121 |
| 8 | 682496 | dll | 0.99990 | 0.99985 | 0.98606 |
| 9 | 860713 | pdf | 0.99990 | 0.99993 | 0.98800 |
| 10 | 988216 | exe | 0.99995 | 0.99995 | 0.97259 |
| 11 | 1395473 | txt | 0.99990 | 0.99996 | 0.99123 |
| 12 | 4472320 | doc | 0.99998 | 0.99995 | 0.98320 |
| 13 | 7820026 | avi | 0.99996 | 0.99996 | 0.99167 |
| 14 | 9227808 | zip | 0.99997 | 0.99998 | 0.99997 |
| 15 | 11580416 | dll | 0.99992 | 0.99998 | 0.99780 |
| 16 | 17486968 | exe | 0.99996 | 0.99997 | 0.99847 |
| 17 | 20951837 | rar | 0.99998 | 0.99996 | 0.99996 |
| 18 | 32683952 | pdf | 0.99997 | 0.99998 | 0.99992 |
| 19 | 44814336 | exe | 0.99991 | 0.99990 | 0.99982 |
| 20 | 62657918 | avi | 0.99997 | 0.99998 | 0.99989 |

Table V
Bit Independence criterion for TDES, AES & PCT algorithms.

|   | Source File Size (bytes) | File type | Bit Independence achieved | | |
|---|---|---|---|---|---|
|   |   |   | TDES | AES | PCT |
| 1 | 330 | dll | 0.49180 | 0.47804 | 0.39186 |
| 2 | 528 | txt | 0.22966 | 0.23056 | 0.20963 |
| 3 | 96317 | txt | 0.41022 | 0.41167 | 0.42861 |
| 4 | 233071 | rar | 0.99899 | 0.99887 | 0.98364 |
| 5 | 354304 | exe | 0.92538 | 0.92414 | 0.93465 |
| 6 | 536387 | zip | 0.99824 | 0.99753 | 0.99265 |
| 7 | 657408 | doc | 0.98111 | 0.98030 | 0.97279 |
| 8 | 682496 | dll | 0.99603 | 0.99560 | 0.96586 |
| 9 | 860713 | pdf | 0.97073 | 0.96298 | 0.96726 |
| 10 | 988216 | exe | 0.91480 | 0.91255 | 0.92929 |
| 11 | 1395473 | txt | 0.25735 | 0.25464 | 0.24621 |
| 12 | 4472320 | doc | 0.98881 | 0.98787 | 0.95390 |
| 13 | 7820026 | avi | 0.98857 | 0.98595 | 0.96813 |
| 14 | 9227808 | zip | 0.99807 | 0.99817 | 0.99804 |
| 15 | 11580416 | dll | 0.86087 | 0.86303 | 0.86049 |
| 16 | 17486968 | exe | 0.83078 | 0.85209 | 0.85506 |
| 17 | 20951837 | rar | 0.99940 | 0.99937 | 0.99936 |
| 18 | 32683952 | pdf | 0.95803 | 0.95850 | 0.95785 |
| 19 | 44814336 | exe | 0.70104 | 0.70688 | 0.82622 |
| 20 | 62657918 | avi | 0.99494 | 0.99451 | 0.99664 |

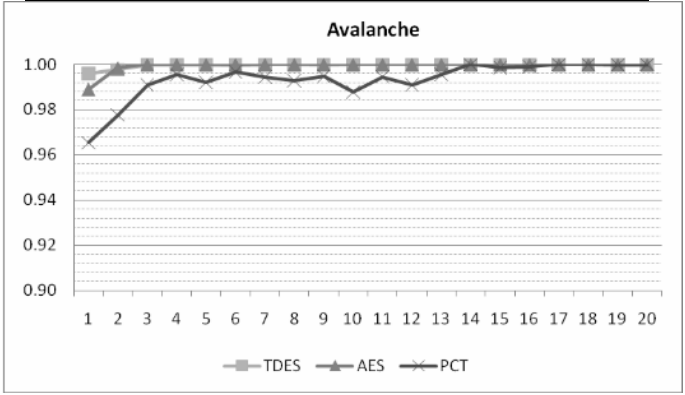

Fig. 3. Comparison of Avalanche effect between TDES, AES and PCT

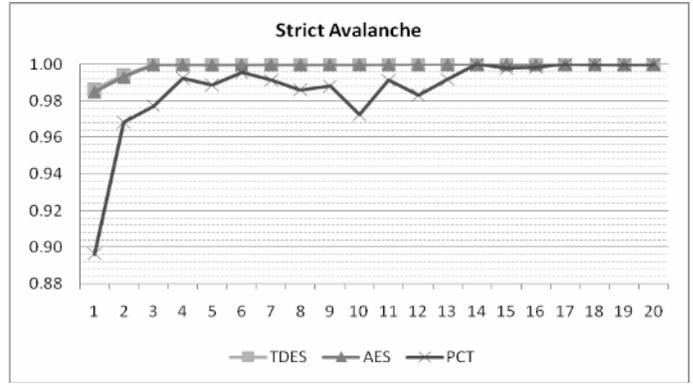

Fig. 4. Comparison of Strict Avalanche effect between TDES, AES and PCT

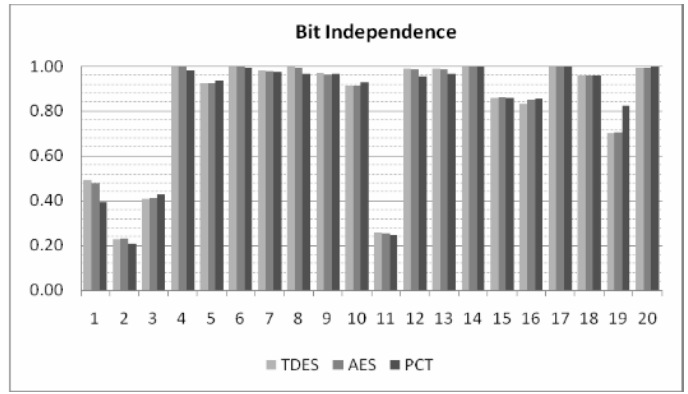

Fig. 5. Comparison of Bit Independence criterion between TDES, AES and PCT

### IV-III. ANALYSIS OF CHARACTER FREQUENCIES

Distribution of character frequencies are analyzed for text file for T-DES, AES and the proposed PCT algorithms. Fig. 9 shows the pictorial representation of distribution of character frequencies for different techniques. Fig. 6(a) shows the distribution of characters in the source file 'redist.txt' (size – 1395473bytes). Fig. 6(b), 6(c) shows the distribution of characters in encrypted files for T-DES and AES respectively. Fig. 6(d) gives the distribution of characters in encrypted file using PCT. All the standard techniques and the proposed PCT technique show a distributed spectrum of characters. From this observation it may be conclude that the proposed technique may obtain good security.

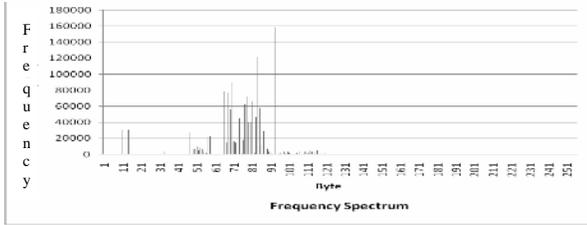

Fig. 6(a). Distribution of characters in source file

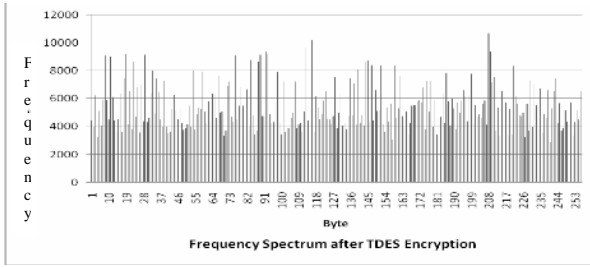

Fig. 6(b): Distribution of characters in TDES

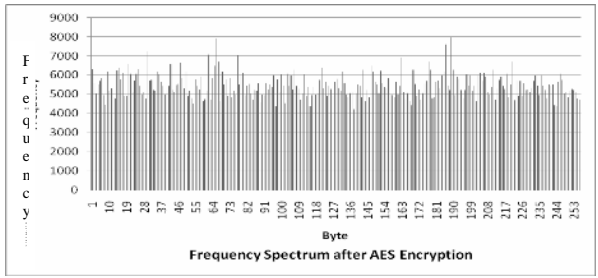

Fig. 6(c). Distribution of characters in AES

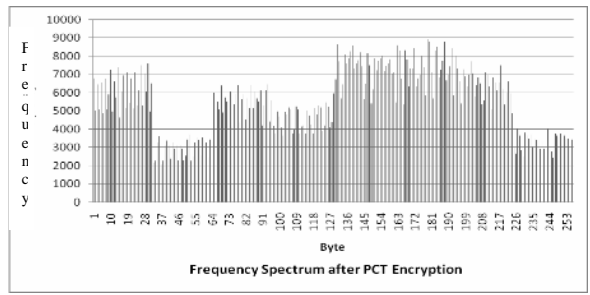

Fig. 6(d). Distribution of characters in PCT

Figure 6 – Frequency Spectrum under different encryption techniques.

## IV-IV. TESTS FOR NON-HOMOGENEITY

The well accepted parametric test has been performed to test the non-homogeneity between source and encrypted files. The large Chi-Square values may confirm the heterogeneity of the source and the encrypted files. The Chi-Square test has been performed using source file and encrypted files for PCT technique and existing TDES and PCT techniques. Table VI shows the values of Chi-Square for different file sizes,.

From Table VI we may conclude that chi-square values depend on the file types as well as file sizes. Text files generally show a large value for chi square. In all cases the chi-square values of the proposed technique is at par with the values from the standard TDES and AES encryptions. The graphical representation of chi-square values are given in Fig. 7 on a logarithmic scale.

Table VI
Chi-Square values

| | Source File Size (in bytes) | File type | ChiSquare | | |
|---|---|---|---|---|---|
| | | | TDES | AES | PCT |
| 1 | 330 | dll | 922.36 | 959.92 | 895.98 |
| 2 | 528 | txt | 1889.35 | 1897.77 | 1940.65 |
| 3 | 96317 | txt | 23492528.45 | 23865067.21 | 20194563.10 |
| 4 | 233071 | rar | 997.78 | 915.96 | 973.19 |
| 5 | 354304 | exe | 353169.83 | 228027.38 | 176908.02 |
| 6 | 536387 | zip | 3279.56 | 3510.12 | 3362.39 |
| 7 | 657408 | doc | 90750.68 | 88706.29 | 88074.14 |
| 8 | 682496 | dll | 29296.79 | 28440.42 | 26668.99 |
| 9 | 860713 | pdf | 59797.35 | 60661.50 | 56190.58 |
| 10 | 988216 | exe | 240186.48 | 245090.50 | 257426.16 |
| 11 | 1395473 | txt | 5833237390.99 | 5545862604.40 | 6778413715.96 |
| 12 | 4472320 | doc | 102678.48 | 102581.31 | 99973.55 |
| 13 | 7820026 | avi | 1869638.73 | 1326136.98 | 808552.25 |
| 14 | 9227808 | zip | 37593.98 | 37424.24 | 36755.77 |
| 15 | 11580416 | dll | 28811486.61 | 17081530.73 | 13773034.65 |
| 16 | 17486968 | exe | 8689664.61 | 8463203.56 | 8003002.32 |
| 17 | 20951837 | rar | 25615.74 | 24785.41 | 26517.84 |
| 18 | 32683952 | pdf | 13896909.50 | 13893011.19 | 15313939.92 |
| 19 | 44814336 | exe | 97756312.18 | 81405043.92 | 500344725.05 |
| 20 | 62657918 | avi | 3570872.51 | 3571648.48 | 4898122.07 |

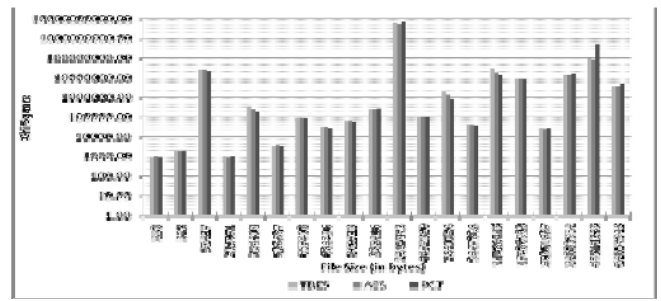

Fig.7. Chi-Square Values for TDES, AES and PCT techniques in logarithmic scale.

## V. CONCLUSION

The PCT Algorithm has been developed keeping "randomness" and "unpredictability" in mind. It essentially works on session key concepts, with a Key and a virtual second key, derived from the main key. Standard language based secure random number generating functions (RFC 1750) (for ex : *class SecureRandom()*[1] in java library ) are based on a certain specific algorithm and formula available for anyone to understand through the open source community – that guarantees generation of unpredictable random numbers giving an extra edge in security for key generation. Still, the PCT algorithm does not totally depend on the random numbers and incorporates a unique key generation algorithm which is partially independent of standard sequence of random - random numbers.

The proposed technique presented in this paper is simple, easy to implement. The key space increases with increase in file size. The major necessities of a good block cipher viz. Avalanche, Strict Avalanche effects and Bit Independence criterion is satisfied and is comparable with industry standards Triple-DES and AES algorithms. The shifting of bits itself gives a high degree of diffusion, and the final XOR with SHA-512 digest of the key gives desired confusion levels.. Performance improves drastically for larger files and is at par or better than AES encryption for very large files and is significantly better than TDES algorithms.

## VI. FUTURE SCOPE

A lot of experiments can be done from the basic shifting of bits, assigning values to MaxIter with corresponding block lengths. For example we may flip even bits and shift odd bits or shift both even and odd bits, one to right and other to left. An increase in the number of iterations, will increase the security and give better results in terms of Avalanche, Strict Avalanche and Bit Independence criteria's, at the cost of increased encryption time. A correct balance between security and time complexity should be the goal.